\documentclass[preprint,showpacs,preprintnumbers,amsmath,amssymb]{revtex4-1}

\usepackage{graphicx}
\usepackage{dcolumn}
\usepackage{bm}

\begin{document}

\title{QCD effective charge and the structure function $F_{2}$ at small-$x$}

\author{E.G.S. Luna,$^{1}$ A.A. Natale,$\, ^{2}$ and A.L. dos Santos$^{1}$}
\affiliation{
$^{1}$Instituto de F\'{\i}sica e Matem\'atica, Universidade Federal de Pelotas, 96010-900, Pelotas, RS, Brazil \\
$^{2}$Instituto de F\'{\i}sica Te\'orica, UNESP - Universidade Estadual Paulista,
Rua Dr. Bento T. Ferraz, 271, Bloco II,
01140-070, S\~ao Paulo - SP,
Brazil}


\begin{abstract}
We study infrared contributions to the QCD description of the HERA data on the structure function $F_{2}$ within
the \textit{generalized} DAS approximation.
We argue that this approximation is a natural one and consistent with the phenomenon of dynamical mass
generation in QCD.
The investigation is performed at next-to-leading order by using the leading-twist expansion
of $F_{2}(x,Q^{2})$ and by adopting an effective charge whose finite infrared behavior is constrained by a
dynamical gluon mass. We propose one ansatz for the behavior of this effective coupling beyond leading order.
The dependence of the experimental data on
the infrared value of the effective charge is used in order to study the asymptotic behavior
of the running gluon mass. The deep inelastic structure function $F_{2}$ obtained
in this approach shows very good agreement with the experimental data.
\end{abstract}

\maketitle

\section{Introduction}

The nucleon structure function $F_2(x,Q^{2})$ at low $Q^{2}$ has been measured in the previously unexplored small-$x$ regime
at the HERA collider. The deep-inelastic scattering of leptons off nucleons is the instrumental tool for high
precision measurements of the quark and gluon content of the nucleons
and the low $Q^{2}$ transition region bring us into a kinematical region where
non-perturbative QCD effects becomes essential in order to understand the proton constitution.
Despite the partonic splitting to be quite well understood through the use of the
Dokshitzer-Gribov-Lipatov-Altarelli-Parisi (DGLAP) evolution equations \cite{dglap}, and these equations being
known to describe the data even at not so large $Q^{2}$, there is no reason to expect that they are reliable in
the very small-$x$ region. However, perturbative QCD effects are expected to become apparent at small-$x$,
where gluon emission off the incoming parton leads to power series in $\alpha_s \ln(1/x)$. Resumming of this series via
the Balitsky-Fadin-Kuraev-Lipatov (BFKL) equation \cite{bfkl}, besides producing an $x^{-\lambda}$ behavior for the
gluon distribution, generates its own characteristic $Q^{2}$ dependence.
Hence approaching the low $Q^{2}$ transition region from the perturbative side makes evident the problem of how to incorporate in an
effective way non-perturbative corrections into the evolution scenario. 

Fortunately, this problem can be properly addressed by bringing up information about the infrared
properties of QCD, more specifically, by considering the possibility that the non-perturbative dynamics of QCD
generate an effective gluon mass at very slow $Q^{2}$ region. This dynamical gluon mass is intrinsically related to an infrared
finite strong coupling constant \cite{agui}, and its existence is strongly supported by recent QCD lattice simulations
\cite{lqcd} as well as by phenomenological results \cite{halzen,luna01,luna02}. 
It is opportune to remember that phenomenological infrared modifications of the strong-coupling constant are quite
usual in the literature \cite{frez,cve}, nevertheless the fact that an infrared finite coupling constant appears as a consequence
of a dynamically generated gluon mass is much less known. Furthermore, the dynamical gluon mass that constrains the finite
coupling constant turns up as the natural infrared cutoff in many perturbative QCD calculations, besides being responsible for a
smooth transition from the perturbative to the non-perturbative QCD behavior \cite{luna01,luna02}.

Our task of calculating infrared contributions to the QCD description of the HERA data on the deep-inelastic structure function,
$F_{2}(x,Q^{2})$, can succeed in a consistent way by analyzing exclusively the small-$x$ region since, in this limit, some of the
existing analytical solutions of the DGLAP equation can be directly used \cite{ball,frichter,yund1,kot3}. Within this
approach the HERA data at small-$x$ is interpreted in terms of the double-asymptotic-scaling (DAS) phenomenon \cite{ball}
related to the asymptotic behavior of the DGLAP equation in asymptotically free field theories \cite{ruj}.
The analytical solutions, valid in principle at very small-$x$ and large-$Q^{2}$ values, can be extended in order to include
the subasymptotic part of the $Q^2$ evolution,
in what is called \textit{generalized} DAS approximation \cite{cve,mank,kot,kot2}, leading to small-$x$ asymptotic predictions
for parton distribution functions evolved from flat $x$ distributions at some starting point $Q_{0}^{2}$ for the DGLAP
evolution. In particular, a recent analysis of $F_2$ and its derivatives
$\partial F_2 /\partial \ln Q^2$ and $\partial \ln F_2/\partial \ln (1/x)$ within this approach shows a good agreement with HERA
data of deep-inelastic scattering for $Q^{2}\gtrsim 1.5$ GeV$^{2}$ \cite{cve}.

One may wonder why the \textit{generalized} DAS approximation works so beautifully in the small-$x$ limit. This fact may be a
signal that the choice of a flat distribution for the parton distribution function $f_a (x,Q^2)$ at some initial value $Q_0^2$ is
quite appropriate for the DGLAP dynamics, more specifically \cite{function},
\begin{eqnarray}
f_a (x,Q^2_0) = A_a  \,\,\,\,\,\,\,\,\, (a=q,g)  \,\, ,
\label{eq1}
\end{eqnarray}
where $A_a$ are unknown constants to be determined from the data, or in other way, that QCD predicts at small-$x$
that $F_{2}(x,Q^{2})$ should exhibit double scaling even at not so large-$Q^{2}$ values, provided only that the small-$x$ behavior
of the partonic distributions at some initial input $Q_{0}^{2}$ is sufficiently soft.
Actually it was also pointed out that a flat gluon distribution in the small-$Q^2$ region appears naturally in
QCD \cite{luna02}, within a model for hadronic cross section including the phenomenon of
dynamical gluon mass generation, which naturally leads to a ``frozen" infrared effective charge. This mechanism, based
on first principles, is probably what is behind the good agreement of the \textit{generalized} DAS approach with the experimental
data.

Hence the purpose of this Letter is to compute the structure function $F_{2}(x,Q^{2})$ of the
proton by means of the {\it generalized} DAS approximation \cite{kot,kot2,cve}, assuming the flat initial parton distributions as
a natural condition for QCD with dynamically generated gluon masses, and compare the results with the experimental
data of $F_{2}(x,Q^{2})$ in the infrared $Q^{2}$ region.
In our calculations the non-perturbative dynamics of QCD is introduced by using the infrared finite QCD effective charge
dependent on the dynamical gluon mass. As this effective strong-coupling has not been determined up to the next-to-leading
order (NLO) approximation, we propose one ansatz for its behavior at higher order.

The Letter is organized as follows: in the next section we introduce the {\it generalized} DAS approach beyond the
leading order (LO),
and discuss the underlying QCD dynamics behind the flat distribution and the frozen effective charge behaviors.
In the Sec. III we propose an ansatz for the NLO behavior of the dynamical strong-coupling based on
the property of multiplicative renormalizability, showing that this effective coupling reproduces the canonical NLO
perturbative behavior at large $Q^{2}$. Our results are presented in the Sec. IV, where the analysis of $F_{2}(x,Q^{2})$ data
is carried on using the formalism developed in the previous sections. In Sec. V we present our conclusions.

\section{The generalized DAS approximation}

The present data of $F_{2}(x,Q^{2})$ imply a steep gluon at small-$x$, and there are some successful descriptions of $F_{2}$ by
means of DGLAP evolution in the NLO approximation \cite{devenish01}. This steep behavior can be generated
from a flat-$x$ gluon distribution at
some initial low $Q_{0}^{2}$ scale, or alternatively it can be directly included into the input distribution to be evolved from
some higher scale.
At sufficiently small-$x$ we must resum the power series in $\alpha_s \ln(1/x)$ via BFKL
equation. The result of this procedure is sensitive to the infrared $k_{T}$ region and, for running $\alpha_s$, it is found that
\begin{eqnarray}
\tilde{g}(x,k_{T}^2) \sim C(k_{T}^2)\, x^{-\lambda} \, ,
\label{g1}
\end{eqnarray}
where $\lambda \sim 0.5$ \cite{martin01}. Here $\tilde{g}(x,k_{T}^{2})$ is the {\it unintegrated} gluon distribution and hence the
resummation program requires knowledge of the gluon for all $k_{T}^2$ including the infrared region. However, in this confinement
region the BFKL equation is not expected to be valid. Ultimately, with decreasing $x$, the singular behavior must be suppressed
by non-perturbative effects.

The problem of calculating these infrared effects can be addressed by the so called QCD-based eikonal
models \cite{luna01,pancheri}, which incorporate soft and semihard processes in the treatment of high energy hadron-hadron
interactions.
At high energies semihard processes are expected to give an increasing and significant part of the total hadronic cross
sections \cite{ryskin01}. Owing to the rapid growing of the number of semihard gluons in the hadron at fixed transverse
momentum, the asymptotic behavior of these cross sections is determined chiefly by the gluon distribution.
In some QCD-based models these small-$x$ semihard gluons play a central role, and a phenomenological
``BFKL-inspired'' gluon distribution is introduced \cite{luna01,luna02},
\begin{eqnarray}
g(x,Q^2) = h(Q^2)\, x^{-J} \, ,
\label{g111}
\end{eqnarray}
which captures all the non-perturbative dynamics via the function $h(Q^2)$. Note that here $g(x,Q^2)$ is the traditional gluon
distribution determined by the parton analysis of the $F_{2}(x,Q^{2})$ data and whose $Q^{2}$ evolution is controlled by
the DGLAP equations, where
\begin{eqnarray}
\tilde{g}(x,k_{T}^2) = \left. \frac{\partial (xg(x,Q^{2}))}{\partial \ln Q^{2}} \right|_{Q^{2}=k_{T}^2} .
\end{eqnarray}

Recently it was shown that, in order to
generate radiatively gluons at small-$x$, a rapid increase of $h(Q^2)$ with the momentum has to be accompanied by a fast
increase of the $J$ in such way that soft values of $J$ are preferred at low $Q^{2}$ \cite{luna02}. This picture
is consistent with the statement that the steeply-rising gluon component is absent at low $Q^{2}$ and as
$Q^{2}$ increases it is generated radiatively through perturbative evolution. In the
Regge-exchange language the quantity $J$, that controls the asymptotic behavior of the total cross sections, is the universal
``soft'' Pomeron intercept, whose value has been phenomenologically determined to be
$J = \alpha_{\mathbb{P}}(0) = 1 + \epsilon \sim 1.1$ \cite{intercept,intercept2}. This is to be contrasted with the ``hard'' or
``Lipatov'' Pomeron intercept $\alpha_{L}(0) = 1 + \lambda \sim 1.5$. Therefore,
fits to a set of hadronic
data through QCD-based models show that $J$ starts at a value where the gluon distribution is almost flat,
$J(Q^{2}\sim 0) \approx \alpha_{\mathbb{P}}(0)$, and as $Q^{2}$ increases the valence-like character of the gluon rapidly disappears
\cite{luna01,luna02}. It is worth noting that these results are corroborated by a MRST analysis of parton distributions
of proton \cite{mrst01}. From fitting the sea quark (S)
and gluon (G) distributions of the default MRST partons to the forms $f_{i}(x,Q^2) = A(Q) x^{-\lambda_{i}(Q^2)}$ as $x\to 0$,
$i=S,G$, they have observed that as $Q^2$
increases from the input scale $Q_{0}^{2}=1$ GeV$^2$ the flat behavior of the gluon rapidly
disappears due to evolution being driven by the much steeper sea. For higher values of $Q^2$ the gluon exponent $\lambda_{G}$
increases rapidly and becomes higher than the sea quark exponent $\lambda_{S}$, since the gluon drives the sea quark via
the $g\to \bar{q}q$ transition. More specifically, $\lambda_{G}$ starts at a value $\lambda_{G} \approx 0$ at
$Q^2 \approx 1$ GeV$^2$, and by $Q^2 \approx 4$ GeV$^2$ it has the value $\lambda_{G}=0.2$.
Hence a flat input gluon distribution at low momenta, that appears as the natural condition for QCD evolution with dynamically
generated gluon masses, also comes out in standard perturbative procedures.

In the {\it generalized} DAS approach \cite{mank,kot,kot2,cve} the subasymptotic corrections are included via the finite parts
of anomalous dimensions of Wilson operators and Wilson coefficients. Remember that in the {\it standard} DAS approximation
only the singular parts of the anomalous dimensions are taken into account. In the {\it generalized} approach the flat input
gluon distribution (\ref{eq1}) determines the small-$x$ asymptotics and, at NLO, the twist-two (leading) term of $F_{2}(x,Q^{2})$
is given by \cite{mank,kot,kot2}
\begin{eqnarray}
F_{2}(x,Q^{2}) = e \left[ f_{q}(x,Q^{2}) + \frac{4T_{R}n_{f}}{3}\,
\frac{\alpha_{s}(Q^{2})}{4\pi}\, f_{g}(x,Q^{2}) \right] \,\, ,
\label{f211}
\end{eqnarray}
where $e=\sum_{i}^{f} e_{i}^{2}/n_{f}$ is the average charge squared of the effective number $n_{f}$ of quarks, $T_{R}=1/2$ is
the color factor for $g\to q\bar{q}$ splitting, and
\begin{eqnarray}
f_{a}(x,Q^{2}) = f_{a}^{+}(x,Q^{2}) + f_{a}^{-}(x,Q^{2})  \,\,\,\,\,\,\,\,\, (a=q,g)  \,\, ;
\label{f222}
\end{eqnarray}
the ``$+$'' and ``$-$'' representation above follows from the solution, at leading twist approximation, of the DGLAP equation
in the Mellin moment space \cite{kot}:
\begin{eqnarray}
f_{a}^{-}(x, Q^{2}) = A_{a}^{-}(Q^{2},Q_{0}^{2})\exp
\left[ -d_{-}(1)s-D_{-}(1)p \right] +{\cal O}(x)  \,\, ,
\label{f233}
\end{eqnarray}
\begin{eqnarray}
f_{g}^{+}(x, Q^{2}) =
A_{g}^{+}(Q^{2},Q_{0}^{2}) \, \tilde{I}_{0}(\sigma) \, \exp \left[
-\bar{d}_{+}(1)s-\bar{D}_{+}(1) p \right] + {\cal O}(\rho)  \,\, ,
\label{f255}
\end{eqnarray}
\begin{eqnarray}
f_{q}^{+}(x, Q^{2}) &=& A_{q}^{+}(Q^{2},Q_{0}^{2}) \left[ \left( 1-\bar{d}_{+-}^{q}(1)\,
\frac{\alpha_{s}(Q^{2})}{4\pi} \right) \rho \, \tilde{I}_{1}(\sigma) +
\frac{20 C_{A}}{3}\, \frac{\alpha_{s}(Q^{2})}{4\pi}
\, \tilde{I}_{0}(\sigma) \right] \nonumber \\
&\times& \exp \left[ -\bar{d}_{+}(1)s - \bar{D}_{+}(1)p \right] + {\cal O}(\rho) \,\, ,
\label{f244}
\end{eqnarray}
where $s = \ln \left[ \alpha_s(Q^2_0 )/\alpha_s(Q^2) \right]$,
$p = \left[ \alpha_s(Q^2_0 ) - \alpha_s(Q^2) \right]/4\pi$,
$D_\pm (n)= d_{\pm\pm} (n)-(\beta_1 /\beta_0) d_\pm (n)$,
$\sigma = 2\sqrt{\left( \hat{d}_{+}s+\hat{D}_{+}p \right) \ln x}$
and
$\rho = \sqrt{ \left( \hat{d}_{+}s+\hat{D}_{+}p \right) /\ln x}
= \sigma /2\ln (1/x)$;
here $\beta_0$ $(\beta_1)$ is the first (second) coefficient of the QCD $\beta$ function, $\tilde{I}_{\nu}$ ($\nu = 0,1$)
are functions related to the modified Bessel function $I_{\nu}$,
and the components of the anomalous
dimension $d_{-}(n)$ as well as of the singular ($\hat{d}$) and regular ($\bar{d}$) parts of
$d_{+}(n) = \hat{d}_{+}/(n-1) + \bar{d}_{+}(n)$, for $n\to 1$, are
\begin{eqnarray}
\hat{d}_{+} = -\frac{4C_{A}}{\beta_{0}}, \hspace{1.3truecm}
\bar{d}_{+}(1) = 1 + \frac{4n_{f}}{3\beta_{0}} \left( 1 - \frac{C_{A}}{C_{F}} \right), \hspace{1.3truecm}
d_{-}(1) = \frac{4C_{F}n_{f}}{3C_{A}\beta_{0}} \,\, ;
\end{eqnarray}
finally, the factors $A_{a}^{+,-}$ and the components of the singular and regular parts of the remaining anomalous dimensions
$D_\pm$ are given by
\begin{eqnarray}
A_{g}^{+}(Q^{2},Q_{0}^{2}) &=& \left[ 1 - \bar{d}_{+-}^{g}(1) \,
  \frac{\alpha_{s}(Q^{2})}{4\pi} \right] A_{g} \nonumber \\
&+& \frac{C_{F}}{C_{A}}\left[ 1 - d_{-+}^{g}(1)\,
  \frac{\alpha_{s}(Q_{0}^{2})}{4\pi}
- \bar{d}_{+-}^{g}(1) \, \frac{\alpha_{s}(Q^{2})}{4\pi} \right] A_{q} \, ,
\end{eqnarray}
\begin{eqnarray}
A_{g}^{-}(Q^{2},Q_{0}^{2}) = A_{g} - A_{g}^{+}(Q^{2},Q_{0}^{2}) \, ,
\end{eqnarray}
\begin{eqnarray}
A_{q}^{+}(Q^{2},Q_{0}^{2}) = \frac{n_{f}}{3C_{A}} \left( A_{g} +
\frac{C_{F}}{C_{A}}\, A_{q} \right) \, ,
\end{eqnarray}
\begin{eqnarray}
A_{q}^{-} (Q^{2},Q_{0}^{2}) = A_{q}- \frac{20C_{A}}{3}\,
\frac{\alpha_{s}(Q_{0}^{2})}{4\pi} \, A_{q}^{+}(Q^{2},Q_{0}^{2}) \, ,
\end{eqnarray}
\begin{eqnarray}
\hat{d}_{++} = \frac{4n_{f}}{9\beta_{0}} \left( 23C_{A}-26C_{F}
\right), \hspace{1.3truecm}
\hat{d}_{+-}^{q} = -\frac{20C_{A}}{3} , \hspace{1.3truecm}
\hat{d}_{+-}^{g} = 0 \, ,
\end{eqnarray}
\begin{eqnarray}
\bar{d}_{++}(1) &=& \frac{8}{3\beta_{0}}\left[
  \frac{C_{A}^{2}}{3}\left(36\zeta(3)+33\zeta(2)
- \frac{1643}{12}\right)\right.\nonumber \\
 &-&\left(2C_{F}\zeta(2)+\frac{43}{9} \, C_{A}-\frac{547}{36} \,
C_{F} + \frac{3}{2}\frac{C_{F}^{2}}{C_{A}}\right)n_{f} \nonumber \\
&-& \left. \frac{13}{18} \frac{C_{F}}{C_{A}} \left(1-2 \,
\frac{C_{F}}{C_{A}}\right)n_{f}^{2}\right] ,
\end{eqnarray}
\begin{eqnarray}
d_{--}(1) &=& \frac{4C_{A}C_{F}}{\beta_{0}}\left(1 - 2 \,
\frac{C_{F}}{C_{A}}\right)\left(2\zeta(3)-3\zeta(2)+\frac{13}{4}
+ \frac{13}{27}\frac{n_{f}^{2}}{C_{A}^{2}}\right) \nonumber \\
&+& \frac{4C_{F}}{3\beta_{0}}\left(4\zeta(2)-\frac{47}{18} + 3 \,
\frac{C_{F}}{C_{A}}\right)n_{f},
\end{eqnarray}
\begin{eqnarray}
\bar{d}_{+-}^{q}(1) =
C_{A}\left(9-3 \, \frac{C_{F}}{C_{A}}-4\zeta(2)\right)-\frac{13}{9}\left(1-2
\, \frac{C_{F}}{C_{A}}\right)n_{f} \, ,
\end{eqnarray}
\begin{eqnarray}
\bar{d}_{+-}^{g}(1) = \frac{20n_{f}}{9} \frac{C_{F}}{C_{A}} , \hspace{1.3truecm}
d_{-+}^{q}(1) = 0 , \hspace{1.3truecm}
d_{-+}^{g}(1) = -\left[C_{A}+\frac{1}{3}\left(1-2\frac{C_{F}}{C_{A}}\right)n_{f}\right] ,
\end{eqnarray}
where $C_{A}=N$, $C_{F}=(N^{2}-1)/2N$, and $\zeta$ is the Riemann zeta function. From now on we set $N=3$ in order to fix
the Casimir color-factors $C_{A}(=3)$ and $C_{F}(=4/3)$. With all these definitions we
can discuss in the next section the QCD effective charge that we shall use in our calculation.

\section{The QCD effective charge at NLO}

Although not extensively known in phenomenological studies, there is increasing evidence that QCD develops an
effective, momentum-dependent mass for the gluons, while preserving the local $SU(3)_{c}$ invariance
of the theory. Of course this mass is not a bare one and at few GeV its signal is already erased from the physical amplitudes,
which merge into the perturbative QCD calculations. In this scenario there is a natural
scale that, in principle, introduces a threshold for gluons to pop up from the vacuum \cite{corn06}. 

Since the gluon mass generation is a purely dynamical effect, the formal tool for tackling this non-perturbative problem,
in the continuum, is provided by the Schwinger-Dyson equations \cite{sde}. These equations constitute an infinite set of coupled
non-linear integral equations governing the dynamics of all QCD Green's functions. In particular, within this framework the
generation of a dynamical gluon mass is associated with the existence of infrared finite solutions for the gluon propagator
$\Delta_{\mu\nu}(q^{2})$ \cite{corn01,corn02,corn03}. In covariant gauges, the gluon propagator has the form
\begin{eqnarray}
\Delta_{\mu\nu}(q^{2}) = -i \left[ P_{\mu\nu}(q)\, \Delta(q^{2}) + \xi \frac{q_{\mu}q_{\nu}}{q^{4}} \right] ,
\end{eqnarray}
where $\xi$ is the gauge-fixing parameter and $P_{\mu\nu}(q) = g_{\mu\nu} - q_{\mu}q_{\nu}/q^{2}$. Infrared finite solutions
($\Delta^{-1}(0) >0$) can be fit by massive propagators on the form $\Delta^{-1}(q^{2}) = q^{2} + m^{2}(q^{2})$, where
$m^{2}(q^{2})$, which depends non-trivially on the momentum transfer $q^{2}$, is the so called dynamical gluon mass.
If the renormalization-group logarithms are included in the Schwinger-Dyson analysis, the non-perturbative generalization
of the QCD running coupling, the effective charge $\bar{\alpha}_{s}(q^{2})$, is
obtained \cite{corn01,corn02}.

Recent studies of a non-linear Schwinger-Dyson equation for the gluon self-energy show that $m^{2}(Q^{2})$ may in fact have two
distinct asymptotic behaviors \cite{agpapa} (note that from now on we adopt the virtuality $Q$ in
our calculations); first, the dynamical gluon mass runs as an inverse power of a logarithm;
second, $m^{2}(Q^{2})$ drops as an inverse power of momentum. The logarithmic running of $m^{2}(Q^{2})$ has been found in previous
studies of linearized Schwinger-Dyson equations to behave as
$m^{2}(Q^{2}) \sim \left( \ln Q^{2}  \right)^{-1-\gamma}$, with $\gamma > 1$ \cite{corn01,corn04}. In the non-linear case \cite{agpapa}
this behavior is rewritten as
\begin{eqnarray}
m^{2}(Q^{2}) = m_{g}^{2} \left[ \frac{\ln \left( \frac{Q^{2}+\rho m_{g}^{2}}{\Lambda^{2}} \right)}{
\ln \left( \frac{\rho m_{g}^{2}}{\Lambda^{2}} \right)} \right]^{-1-\gamma_{1}} ,
\label{eqlog}
\end{eqnarray}
where $\gamma_{1} = -6(1+c_{2}-c_{1})/5$; here $c_{1}$ and $c_{2}$ are parameters of the ansatz for the
(fully dressed) three-gluon vertex used in the numerical analysis of the gluon self-energy. Their values are restricted by
a ``mass condition'' which controls the behavior of the dynamical mass in the ultraviolet region. In the case of a
logarithmic running, $c_{1} \in [0.15,0.4]$ and $c_{2} \in [-1.07,-0.92]$ \cite{agpapa}; the values of the parameters
$\rho$ and $m_{g}$, which control the behavior of $m^{2}(Q^{2})$ in the infrared region, are also restricted by the mass condition,
and general constraints are satisfied for $\rho \in [1.0, 8.0]$ and $m_{g} \in [300, 800]$ MeV \cite{private}.
It is worth mentioning that the dynamical gluon mass was found for the first time by Cornwall to be equal to \cite{corn01}
\begin{eqnarray} 
m^2(Q^2)=m_g^2  \left[ \frac{\ln \left( \frac{Q^2 + 4
   m_g^2}{\Lambda^2} \right)}{\ln \left( \frac{4 m_g^2}{\Lambda^2}
   \right)} \right]^{- \frac{12}{11}},
\label{eqmas}
\end{eqnarray} 
where the infrared mass value $m_g$ is phenomenologically determined and typically of the order
$m_g = 500 \pm 200$ MeV \cite{corn01,halzen,luna01,luna02,nat}. Note that the Cornwall expression (\ref{eqmas}) is a special
case of the logarithmic running mass (\ref{eqlog}); more specifically, (\ref{eqmas}) can be obtained from (\ref{eqlog})
by fixing $\rho=4$ and $\gamma_{1} = 1/11$.

A power-law running behavior for $m^{2}(Q^{2})$ was first envisaged in \cite{corn01,corn05}.
According to an OPE calculation the most probable asymptotic behavior of the running gluon mass is proportional
to $1/Q^2$ \cite{lav}. At the level of an non-linear Schwinger-Dyson equation this asymptotic behavior is given by 
\begin{eqnarray}
m^{2}(Q^{2}) = \frac{m_{g}^{4}}{Q^{2}+m_{g}^{2}} \left[ \frac{\ln \left( \frac{Q^{2}+  \rho m_{g}^{2}}{\Lambda^{2}} \right)}{
\ln \left( \frac{\rho m_{g}^{2}}{\Lambda^{2}} \right)} \right]^{\gamma_{2}-1}  \, ,
\label{eqpo}
\end{eqnarray}
where $\gamma_{2} = (4 + 6 c_{1})/5$; for power law running the mass condition imposes $c_{1} \in [0.7,1.3]$; the $\rho$ and
$m_{g}$ parameters are constrained to lie in the same interval as before, namely $\rho \in [1.0, 8.0]$ and
$m_{g} \in [300, 800]$ MeV \cite{agpapa,private}.

The results of Ref.\cite{agpapa} are precise with respect to the gross asymptotic behavior of the running gluon
mass which are represented by our equations (21) and (23). Eq.(5.2) of that reference contains a broad definition
of the mass function used to match the numerical results, and it must be said that the approximations in there
exclude the ghost fields and the regularization of quadratic divergences is obtained through what is called
``tadpole-condition". This procedure does not determine $\bar{\alpha}_s (0)$ and $m_g$ univocally, and leads to a
dispersion on the frozen coupling infrared behavior. However, this uncertainty is
systematically reduced by the aforementioned phenomenological studies \cite{halzen,luna01,luna02}, which determine a frozen
value for the LO effective charge of the order $\bar{\alpha}_{s}(0) \sim 0.7 \pm 0.2$. Our result for the NLO frozen behavior,
$\bar{\alpha}^{NLO}_{s}(0) \sim 0.6$, despite not being directly comparable to LO results, gives support to the
statement that the dynamical gluon mass $m_{g}$ is not strongly dependent on the perturbation order.
Moreover, our value $\bar{\alpha}^{NLO}_{s}(0) \sim 0.6$ is totally consistent with the frozen value
$\alpha_{s}(0)/\pi \sim 0.19$, obtained very recently from an analytic QCD model \cite{valenzuela}.

Given the running behavior of $m^{2}(Q^{2})$, the leading-order QCD effective charge $\bar{\alpha}_{s}(Q^2)$ is written as
\begin{eqnarray}
\bar{\alpha}_{s} (Q^2) =
\frac{1}{b_{0} \ln \left( \frac{Q^2 + 4m^2(Q^2)}{\Lambda^2} \right) } ,
\label{eq31}
\end{eqnarray}
where $b_{0} = \beta_{0}/4\pi = (1/4\pi)[(11C_{A} - 2n_f)/3 ]$ and $\Lambda \equiv \Lambda_{QCD}^{LO}$. The effective charge
clearly shows the existence of an infrared fixed point as $Q^2\rightarrow  0$, i.e., the dynamical mass term tames the Landau
pole and $\bar{\alpha}_{s}$ freezes at a finite value in the infrared limit. It must be stressed that the fixed point does not
depend on a specific process, it is uniquely obtained as we fix $\Lambda$ and, in principle, it should be  exactly determined if
we knew how to solve QCD. Note that in the limit $Q^{2} \gg \Lambda^{2}$ the dynamical mass $m(Q^{2})$ vanishes, and the effective
charge (\ref{eq31}) matches with the one-loop perturbative QCD coupling $\alpha_{s}(Q^{2})$. It means that the asymptotic
ultraviolet behavior of the LO running coupling, obtained from the renormalization group equation perturbation theory,
\begin{eqnarray}
\alpha_{s}^{LO} (Q^2 \gg \Lambda^2) \sim
\frac{1}{b_{0} \ln \left( \frac{Q^2 }{\Lambda^2} \right) } \,\, ,
\label{eq41}
\end{eqnarray}
is reproduced in solutions of Schwinger-Dyson equations, provided only that the truncation method employed
in the analysis preserves the multiplicative renormalizability (MR). Since the MR is an important feature of gauge field theories,
and holds for any renormalization scale, we argue that a QCD effective charge at NLO, $\bar{\alpha}_{s}^{NLO}$, can be successfully
built by saturating the two-loop perturbative strong coupling $\alpha_{s}^{NLO}$, that is, by introducing the replacement
$\alpha_{s}^{NLO}(Q^{2}) \to \bar{\alpha}_{s}^{NLO}(Q^{2}) = \alpha_{s}^{NLO}(Q^{2} + 4m^{2}(Q^{2}))$ into the perturbative result. In
this way, the QCD effective charge at NLO is given by
\begin{eqnarray}
\bar{\alpha}_{s}^{NLO}(Q^{2}) = \frac{1}{b_{0}\ln\left(\frac{Q^{2} +
4m^{2}(Q^{2})}{\Lambda^{2}}\right)}\left[1-\frac{b_{1}}{b_{0}^{2}}\frac{\ln\left(\ln\left(\frac{Q^{2} +
4m^{2}(Q^{2})}{\Lambda^{2}}\right)\right)}{\ln\left(\frac{Q^{2} + 4m^{2}(Q^{2})}{\Lambda^{2}}\right)} \right],
\label{ansatz2}
\end{eqnarray}
where $b_{1} = \beta_{1}/16\pi^{2} = (1/16\pi^{2})[(34C_{A}^{2} - n_f(10C_{A}+6C_{F}))/3 ]$ and $\Lambda = \Lambda_{QCD}^{NLO}$.
Note that in the limit $Q^{2} \gg \Lambda^{2}$ the effective charge (\ref{ansatz2}) matches with the canonical two-loop
perturbative coupling, ${\alpha}_{s}^{NLO}$, in such a way that the relation
\begin{eqnarray}
\frac{{\bar{\alpha}}_{s}^{NLO}}{\bar{\alpha}_{s}^{LO}} = \frac{{\alpha}_{s}^{NLO}}{{\alpha}_{s}^{LO}} 
\label{eqan}
\end{eqnarray}
is valid in the ultraviolet region. This relation is expected to be valid if the Schwinger-Dyson equation is renormalized
multiplicatively.

\begin{figure}
\vspace{1.0cm}
\begin{center}
\includegraphics[height=.70\textheight]{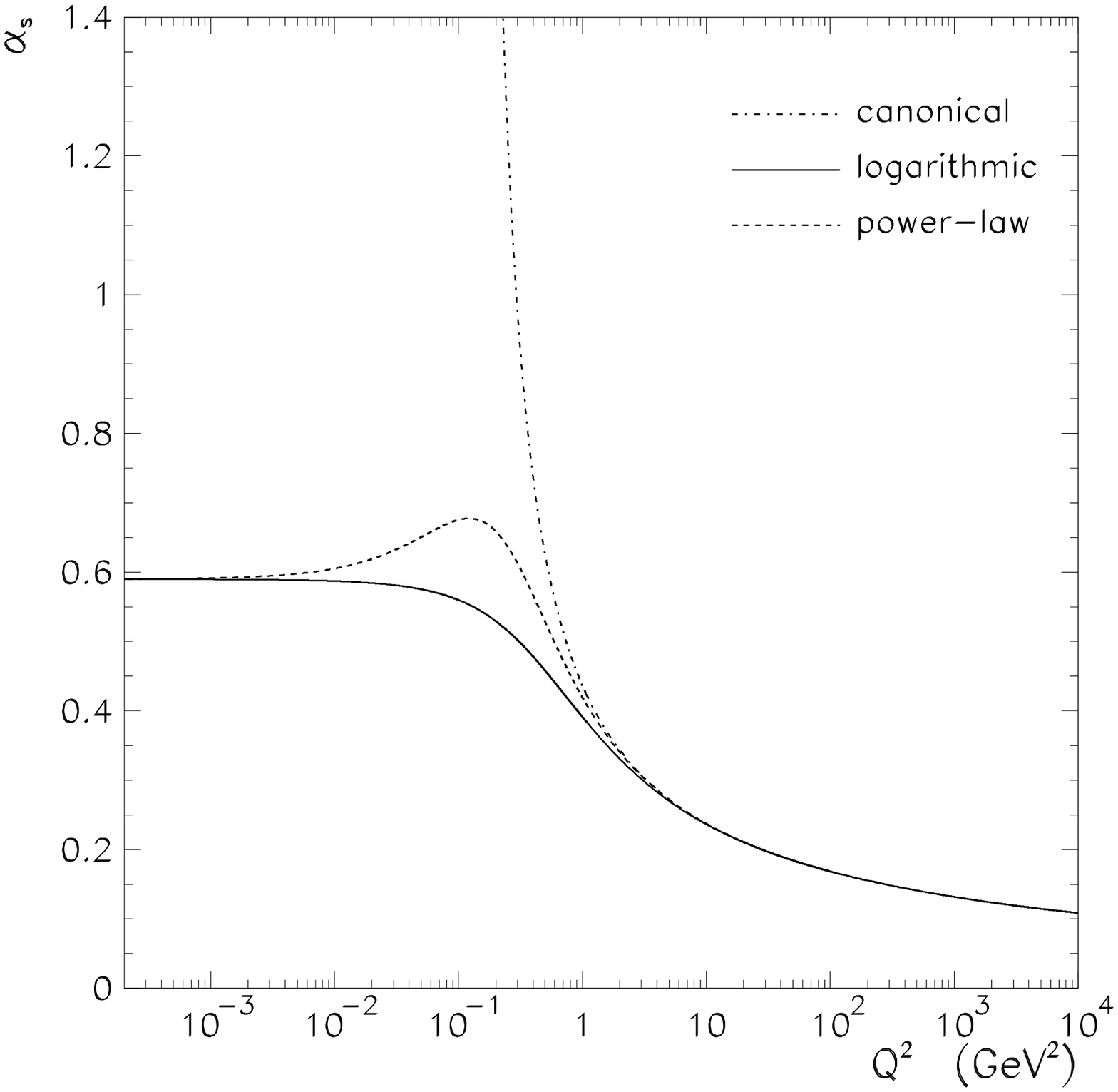}
\caption{The canonical (perturbative) coupling constant and the QCD effective charge with logarithmic and power-law mass running at NLO.}
\label{fig1}
\end{center}
\end{figure}

We have created other ansatzes preserving the relation (\ref{eqan}), where an intermediate scale was introduced in order to
separate the perturbative and non-perturbative regions, but they did not introduce significant differences in the behavior of
$\bar{\alpha}_{s}^{NLO}$ in the infrared region. Thus we have adopted the coupling (\ref{ansatz2}) as the standard ansatz in our
calculations of the structure function $F_{2}(x,Q^{2})$ of the proton. We do not expect that any phenomenological calculation
using the logarithmic and power-law running dynamical masses will be strongly dependent on the asymptotic behavior as they are
on the infrared one. Therefore, since the calculation is quite dependent on the behavior of the effective charge in
the infrared region, and our analysis includes HERA data sets at low and moderate $Q^{2}$, we carry out two independent
global fits to HERA data: in the first one we adopt the effective charge (\ref{ansatz2}) with a logarithmic mass running,
expression (\ref{eqlog}); in the second fit the effective charge runs through the power-law mass running, expression (\ref{eqpo}).
Within this procedure we can investigate if the experimental data can differentiate these solutions.
The different momentum behaviors of the canonical (perturbative) and the effective charges are shown in Fig.\ref{fig1}.

\section{Results}

From the formalism discussed in the previous sections,
we analyze $F_{2}(x,Q^{2})$ data sets at low and moderate $Q^{2}$ values \cite{heradata}, by
adding the statistic and systematic errors in quadrature. We carry out global fits to $F_{2}$ data by means of a $\chi^{2}$
fitting procedure with an interval $\chi^{2}-\chi^{2}_{min}$ corresponding to the projection of the $\chi^{2}$ hypersurface
containing 90\% of probability. 
To keep the analysis as simple as possible, we fix $n_{f}=4$ and $\Lambda = 284$ GeV. These choices are not only consistent
to NLO procedures, but are also the same ones adopted in Ref. \cite{cve}. Concerning the QCD effective charges, we fix in all the
fits $\rho=4$, since this is the optimal value obtained by Cornwall in order to reproduce the numerical results of a gauge
invariant Schwinger-Dyson equation for the gluon propagator \cite{corn01}. Moreover, we observe that the gluon mass scale
$m_{g}$ is not very sensitive to $\rho$ (at least in this leading-twist operator analysis), changing by about 12\% (4\%) when
$\rho$ ranges from 1.0 to 8.0 in the case of a logarithmic (power-law) running behavior.

\begin{figure}
\vspace{1.0cm}
\begin{center}
\includegraphics[height=.70\textheight]{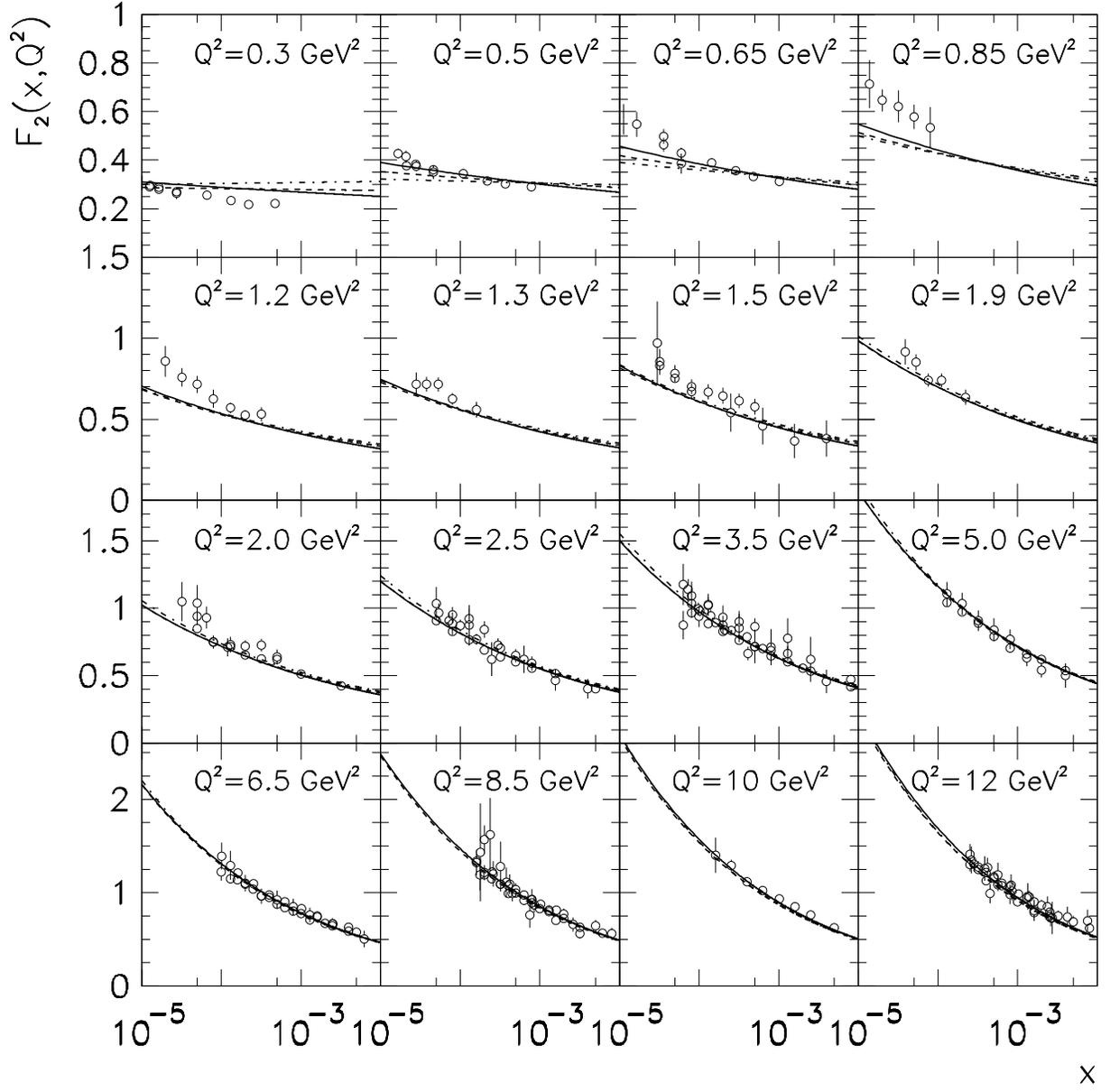}
\caption{Fits of the $x$ dependence of
$F_2 (x, Q^2)$ for specific $Q^2$ with the QCD effective charge dependent on a dynamical gluon mass. The solid curves were obtained
considering the logarithmic running gluon mass, Eq.(\ref{eqlog}), whereas the dashed curves correspond to the power-law running case,
Eq.(\ref{eqpo}). The dotted-dashed curves correspond to the perturbative behavior at NLO. }
\label{fig2}
\end{center}
\end{figure}

Our first analysis consisted in the determination of the $A_g$, $A_q$ and $Q_0^2$ values from a global fit to $F_{2}$ data
using the canonical (perturbative) QCD coupling at NLO. The $\chi^{2}/DoF$ for this fit was 2.88. These values are shown in
Table \ref{tab:tabl01}, and the structure functions corresponding to these values are shown by the dotted-dashed
curves in Fig.\ref{fig2}. It is clear from the relatively high value of $\chi^{2}/DoF$ obtained in this fit, as well as from the
curves depicted in Fig.\ref{fig2}, that the canonical version of the coupling constant provides a worse fit for the $x$
dependence of $F_{2} (x, Q^2)$ for specific $Q^2$ bins, and the disagreement is larger for smaller values of $Q^2$. This result
was expected in the light of an earlier analysis on $F_{2}$ data, which showed the requirement for theoretical improvements in the
QCD canonical coupling for smaller $Q^{2}$ values \cite{cve}. More specifically, it was shown that modifications in the
strong coupling based on the K\"all\'en-Lehmann $Q^{2}$ analyticity \cite{shirkov01}, or on a purely phenomenological freezing,
improve the description of the experimental $F_{2}$ data at low $Q^{2}$. Our results for $F_2 (x,Q^2)$ in the case of the
standard perturbative coupling are therefore similar to the ones of Ref.\cite{cve}. It is important to stress that the analytic
coupling constant used in the Ref.\cite{cve} has a frozen value
$\alpha_{s}(0)\simeq 1.398$, which is not consistent with phenomenological results using the gluon dynamical mass.
Moreover, this analytic coupling, as pointed out by Cveti\v{c}, K\"ogerler and Valenzuela \cite{valenzuela}, does
not give the correct value of the well-measured semihadronic $\tau$ decay ratio $r_{\tau}$, namely
$r_{\tau}^{exp}=0.203\pm 0.004$. However, the analytic coupling version of the Ref.\cite{valenzuela}, besides
generating a frozen value similar to the one obtained in this Letter, reproduces successfully the experimental value
of $r_{\tau}$. 

In the second analysis we carried out a global fit to $F_{2}$ data using the QCD effective charge with a logarithmic running mass,
namely, using the expressions (\ref{eqlog}) and (\ref{ansatz2}). In principle, the infrared mass scale $m_{g}$ and
the factor $\gamma_{1}$ are also, together with $A_g$, $A_q$ and $Q_0^2$, fitting parameters to be determined, where the
constraints on $c_{1}$ and $c_{2}$ are satisfied for $\gamma_{1} \in [0.084, 0.564]$. However, the best $\chi^{2}$ value for this fit
is obtained at $\gamma_{1} = 0.084$. If we carry out global fits to different combinations of $F_{2}$ data sets, as for example taking
into account only data sets with $Q^{2} < 2.0$ GeV$^{2}$ or $Q^{2} \ge 2.0$, we observe that the optimum $\chi^{2}$ for each fit has all
$\gamma_{1} = 0.084$. Therefore, we have set $\gamma_{1} = 0.084$ in our subsequent analysis. The $\chi^{2}/DoF$ obtained by this fit
was 1.87. The values of $m_{g}$, $A_g$, $A_q$ and $Q_0^2$ are show in the second line of the Table \ref{tab:tabl01}. The structure
functions corresponding to these values are shown by the solid curves in Fig.\ref{fig2}.

In the sequence we have obtained $m_{g}$, $A_g$, $A_q$ and $Q_0^2$ by means of a fit to $F_{2}$ data using the QCD effective charge
with a power-law running mass, namely, using the expressions (\ref{eqpo}) and (\ref{ansatz2}). In this analysis $\gamma_{2}$,
constrained by $c_{2}$ to lie in the interval $[1.64, 2.36]$, is set to $\gamma_{2} = 2.36$, since the optimum $\chi^{2}$ for fits
to different combinations of data is obtained for $\gamma_{2} = 2.36$. The $\chi^{2}/DoF$ for this global fit
was 2.13. The values of $m_{g}$, $A_g$, $A_q$ and $Q_0^2$ are show in the third line of the Table \ref{tab:tabl01}. The theoretical
$F_{2}$ results corresponding to these values are shown by the dashed curves in Fig.\ref{fig2}.

Note that all $Q_0^2$ dependence enters into the $\alpha_s$ definition. In our approach $Q_0^2$ will always appear added to a
factor $4m^2(Q^{2}_{0})$. Particularly in the case of log-running gluon mass the $Q^2_0$ dependence is not felt strongly, as
a consequence of a very flat infrared coupling constant. In this case the ``effective'' initial scale
$Q^{2}_{eff,0}$ is dominated by the term $4m^2(Q^{2}_{0})$, namely
$Q^{2}_{eff,0}=0.009 + 4m^{2}(0.009) \approx 0.534$ GeV$^{2}$.
On the other hand a power-law running gluon mass leads to a $\alpha_s$ behavior that matches very fast with the
standard perturbative one and consequently a stronger dependence on the $Q^2_0$ parameter, where
$Q^{2}_{eff,0}=0.029 + 4m^{2}(0.029) \approx 0.456$ GeV$^{2}$. These values are close to the result from Ref.\cite{cve}
for the NLO fit using a frozen coupling, namely $Q^{2}_{0}=0.589\pm 0.006$ GeV$^{2}$.

It seems that owing to the larger $\chi^{2}/DoF$ value obtained in the power-law mass analysis the data favors the logarithmic
running mass case, but it must be said that the differences are quite subtle and in both cases we have a substantial
improvement in the agreement of the theory with the experimental HERA data at low $Q^{2}$.

\section{Conclusions}

In this Letter we have computed the structure function $F_{2}(x,Q^{2})$ of the proton by means of the {\it generalized} DAS
approximation with a QCD effective charge at NLO. This effective strong coupling is finite in the infrared region and
naturally connected to the phenomenon of dynamical gluon mass generation in QCD. Its basically flat behavior below the $m_g$
scale indicates the existence of an infrared fixed point.

We have observed that an infrared finite coupling constant is fundamental in order to improve the description of the $F_{2}(x,Q^{2})$
experimental data at low $Q^{2}$, what can easily be seen if we compare the results using effective charges with the one using the
canonical perturbative coupling. Such fact was already observed in Ref.\cite{cve}, in which a phenomenological infrared finite
coupling was also adopted. We stress that our QCD effective charges, despite the same freezing behavior of the couplings adopted
in \cite{cve}, are obtained from the QCD Lagrangian, i.e., they are derived from first principles, in a scenario where infrared
effects are taken into account.
Moreover, we have argued that a flat $x$ initial condition in the DGLAP evolution equations, which determine the basic role of the
singular parts of the anomalous dimensions in the generalized DAS approach, is naturally related to what is expected from
non-perturbative QCD, i.e., from QCD with dynamically generated gluon masses. 

Through global fits to $F_{2}$ data we have obtained the best values of the infrared mass scale in the case of the logarithmic
and power-law running mass, $m_{g}=364\pm 26$ MeV and $m_{g}=355\pm 27$ MeV, respectively. It is important noting that these infrared
scale values are of the same order of magnitude as the values obtained in other calculations of strongly interacting processes
\cite{halzen,luna01,luna02}. These results corroborate theoretical analyzes considering the generation of a dynamical gluon
mass in non-perturbative QCD.

Our results show that the leading-twist approximation of the Wilson operator product expansion is quite accurate on the
description of the structure function $F_{2}$ data. However, our results using QCD effective charges indicates that, at principle,
in order to differentiate the logarithmic running mass from the power-law one, a higher-twist study is necessary. An analysis using
higher twist corrections to the expansion of $F_{2}(x,Q^{2})$ is in progress.

\section*{Acknowledgments}
We are grateful to A.C. Aguilar, V.P. Gon\c{c}alves and W.K. Sauter for valuable discussions. This research was partially
supported by the Conselho Nacional de Desenvolvimento Cient\'{\i}fico e Tecnol\'ogico (CNPq) and by the Coordena\c{c}\~ao de
Aperfei\c{c}oamento de Pessoal de N\'{\i}vel Superior (CAPES).

\newpage

\begin{table}
\centerline{%
\begin{tabular}{|l|c|c|c|c|c|}
\hline \hline
\, Coupling   &  $m_{g}$ [MeV] &   $A_{g}$            &   $A_{q}$           & $Q_{0}^{2}$ [GeV$^{2}$]& $\chi^{2}/DoF$ \\
\hline
\, Canonical   &      -       &  -0.339$\pm$0.019  &  1.119$\pm$0.025  &  0.414$\pm$0.016  &  2.88 \\
\, Logarithmic &  364$\pm$26  &  -0.084$\pm$0.063  &  0.843$\pm$0.069  &  0.009$\pm$0.116  &  1.87 \\
\, Power-Law   &  355$\pm$27  &  -0.253$\pm$0.041  &  1.018$\pm$0.565  &  0.029$\pm$0.008  &  2.13 \\
\hline
\hline
\end{tabular}}
\caption{Values of the parameters $m_{g}$, $A_g$, $A_q$ and $Q^2_0$ resulting from the global fit to $F_{2}$ data.
The errors were obtained assuming a confidence region of the parameters of 90\%.}
\label{tab:tabl01}
\end{table}




\begin{thebibliography}{99}

\bibitem{dglap} V.N. Gribov, L.N. Lipatov, Sov. J. Nucl. Phys. 15 (1972) 438 [Yad. Fiz. 15 (1972) 781];
L.N. Lipatov, Sov. J. Nucl. Phys. 20 (1975) 94 [Yad. Fiz. 20 (1974) 181];
G. Altarelli, G. Parisi, Nucl. Phys. B 126 (1977) 298;
Yu.L. Dokshitzer, Sov. Phys. JETP 46 (1977) 641 [Zh. Eksp. Teor. Fiz. 73 (1977) 1216].

\bibitem{bfkl} L.N. Lipatov, Sov. J. Nucl. Phys. 23 (1976) 338;
V.S. Fadin, E.A. Kuraev, L.N. Lipatov, Sov. Phys. JETP 44 (1976) 443;
V.S. Fadin, E.A. Kuraev, L.N. Lipatov, Sov. Phys. JETP 45 (1977) 199;
Y.Y. Balitsky, L.N. Lipatov, Sov. J. Nucl. Phys. 28 (1978) 822.

\bibitem{agui} A. C. Aguilar, A. A. Natale, P. S. Rodrigues da Silva, Phys. Rev. Lett. 90 (2003) 152001.

\bibitem{lqcd} F.D.R. Bonnet, et al., Phys. Rev. D 64 (2001) 034501;
A. Cucchieri, T. Mendes, A. Taurines, Phys. Rev. D 67 (2003) 091502(R);
P.O. Bowman, et al., Phys. Rev. D 70 (2004) 034509;
A. Sternbeck, E.-M. Ilgenfritz, M. Muller-Preussker, A. Schiller, Phys. Rev. D 72 (2005) 014507;
A. Sternbeck, E.-M. Ilgenfritz, M. Muller-Preussker, Phys. Rev. D 73 (2006) 014502;
Ph. Boucaud, et al., JHEP 0606 (2006) 001;
P.O. Bowman, et al., hep-lat/0703022;
I.L. Bogolubsky, E.M. Ilgenfritz, M. Muller-Preussker, A. Sternbeck, Phys. Lett 676 (2009) 69;
O. Oliveira, P. J. Silva, arXiv:0910.2897 [hep-lat];
O. Oliveira, P. J. Silva, arXiv:0911.1643 [hep-lat];
A. Cucchieri, T. Mendes, E.M.S. Santos, Phys. Rev. Lett. 103 (209) 141602;
A. Cucchieri, T. Mendes, Phys. Rev. D 81 (2010) 016005;
D. Dudal, O. Oliveira, N. Vandersickel, Phys. Rev. D 81 (2010) 074505.

\bibitem{halzen} F. Halzen, G. Krein, A.A. Natale, Phys. Rev. D 47 (1993) 295;
M.B. Gay Ducati, F. Halzen, A.A. Natale, Phys. Rev. D 48 (1993) 2324;
A.C. Aguilar, A. Mihara, A.A. Natale, Phys. Rev. D 65 (2002) 054011;
M.B. Gay Ducati, W.K. Sauter, Phys. Lett. B 521 (2001) 259;
M.B. Gay Ducati, W.K. Sauter, Phys. Rev. D 67 (2003) 014014;
E.G.S. Luna, Phys. Lett. B 641 (2006) 171;
E.G.S. Luna, A.A. Natale, Phys. Rev. D 73 (2006) 074019;
E.G.S. Luna, Braz. J. Phys. 37 (2007) 84;
E.G.S. Luna, in: AIP Conference Proceedings, vol. 1296, American Institute of Physics, New York, 2010, p. 183;
E.G.S. Luna, A.L. dos Santos, in: AIP Conference Proceedings, vol. 1296, American Institute of Physics, New York, 2010, p. 330.

\bibitem{luna01} E.G.S. Luna, A.F. Martini, M.J. Menon, A. Mihara, A.A. Natale, Phys. Rev. D 72 (2005) 034019.

\bibitem{luna02} E.G.S. Luna, A.A. Natale, C.M. Zanetti, Int. J. Mod. Phys. A 23 (2008) 151.

\bibitem{frez} E. Eichten, et al., Phys. Rev. Lett. 34 (1975) 369;
E. Eichten, et al., Phys. Rev. D 21 (1980) 203;
J.L. Richardson, Phys. Lett. B 82 (1979) 272;
G. Parisi, R. Petronzio, Phys. Lett. B 94 (1980) 51;
T. Barnes, F. E. Close, S. Monaghan, Nucl. Phys. B 198 (1982) 380;
S. Godfrey, N. Isgur, Phys. Rev. D 32 (1985) 189;
A.C. Mattingly, P. M. Stevenson, Phys. Rev. Lett. 69 (1992) 1320;
A.C. Mattingly, P. M. Stevenson, Phys. Rev. D 49 (1994) 437;
Yu.L. Dokshitzer, B.R. Webber, Phys. Lett. B 352 (1995) 451;
Yu.L. Dokshitzer, G. Marchesini, B.R. Webber, Nucl. Phys. B 469 (1996) 93;
M. Anselmino, F. Murgia, Phys. Rev. D 53 (1996) 5314;
B. Badelek, J. Kwiecinsky, A. Stasto, Z. Phys. C 74 (1997) 297;
A. Mihara, A.A. Natale, Phys. Lett. B 482 (2000) 378.

\bibitem{cve} G. Cveti\v{c}, A.Y. Illarionov, B.A. Kniehl, A.V. Kotikov, Phys. Lett. B 679 (2009) 350.

\bibitem{ball} R.D. Ball, S. Forte, Phys. Lett. B 336 (1994) 77;
R.D. Ball, S. Forte, Acta Phys. Polon. B 26 (1995) 2097;
R.D. Ball, S. Forte, Nucl. Phys. B (Proc. Suppl.) 54A (1997) 163.

\bibitem{frichter} G.M. Frichter, D.W. McKay, J.P. Ralston, Phys. Rev. Lett. 74 (1995) 1508.

\bibitem{yund1} C. L\'opes, F. Barreiro, F.J. Yndur\'ain, Z. Phys. C 72 (1996) 561;
K. Adel, F. Barreiro, F.J. Yndur\'ain, Nucl. Phys. B 495 (1997) 221.

\bibitem{kot3} A.V. Kotikov, Mod. Phys. Lett. A 11 (1996) 103;
A.V. Kotikov, Phys. Atom. Nucl. 59 (1996) 2137 [Yad. Fiz. {\bf 59} (1996) 2219].

\bibitem{ruj} A. De R\'ujula, S.L. Glashow, H.D. Politzer, S.B. Treiman, F. Wilczek, A. Zee, Phys. Rev. D 10 (1974) 1649.

\bibitem{mank} L. Mankiewicz, A. Saalfed, T. Weigl, Phys. Lett. B 393 (1997) 175.

\bibitem{kot} A.V. Kotikov, G. Parente, Nucl. Phys. B 549 (1999) 242.

\bibitem{kot2} A.Y. Illarionov, A. V. Kotikov, G. Parente Bermudez, Phys. Part. Nucl. 39 (2008) 307.

\bibitem{function} Note that in the expression (\ref{eq1}) the parton distributions $f_a (x,Q^2_0)$ are multiplied
by $x$, namely, $f_{g}(x,Q^{2}) \equiv x g(x,Q^{2})$, $f_{q}(x,Q^{2}) \equiv x q(x,Q^{2})$.

\bibitem{devenish01} A.M. Cooper-Sarkar, R.C.E. Devenish, A. de Roeck, Int. J. Mod. Phys. A 13 (1998) 3385;
J. Breitweg, et al., Eur. Phys. J. C 7 (1999) 609;
C. Adloff, et al., Eur. Phys. J. C 13 (2000) 609.

\bibitem{martin01} A.J. Askew, J. Kwiecinski, A.D. Martin, P.J. Sutton, Phys. Rev. D 47 (1993) 3775;
A.J. Askew, J. Kwiecinski, A.D. Martin, P.J. Sutton, Phys. Rev. D 49 (1994) 4402.

\bibitem{pancheri} R.M. Godbole, A. Grau, G. Pancheri, Y.N. Srivastava, arXiv:1001.4749.

\bibitem{ryskin01} L.V. Gribov, E.M. Levin, M.G. Ryskin, Phys. Rep. 100 (1983) 1;
E.M. Levin, M.G. Ryskin, Phys. Rep. 189 (1990) 267.

\bibitem{intercept} A. Donnachie, P.V. Landshoff, Phys. Lett. B 296 (1992) 227;
R.J.M. Covolan, J. Montanha, K. Goulianos, Phys. Lett. B 389 (1996) 176;
J.R. Cudell, K. Kang, S.K. Kim, Phys. Lett. B 395 (1997) 311;
E.G.S. Luna, M.J. Menon, Phys. Lett. B 565 (2003) 123;
R.F. Avila, E.G.S. Luna, M.J. Menon, Phys. Rev. D 67 (2003) 054020; 
E.G.S. Luna, M.J. Menon, J. Montanha, Nucl. Phys. A 745 (2004) 104;
E.G.S. Luna, M.J. Menon, J. Montanha, Braz. J. Phys. 34 (2004) 268. 

\bibitem{intercept2} E.G.S. Luna, V.A. Khoze, A.D. Martin, M.G. Ryskin, Eur. Phys. J. C 59 (2009) 1;
E.G.S. Luna, V.A. Khoze, A.D. Martin, M.G. Ryskin, Eur. Phys. J. C 69 (2010) 95.

\bibitem{mrst01}  A.D. Martin, R.G. Roberts, W.J. Stirling, R.S. Thorne, Eur. Phys. J. C 4 (1998) 463.

\bibitem{corn06} J.M. Cornwall, A. Soni, Phys. Lett. B 120 (1983) 431;
J.M. Cornwall, A. Soni, Phys. Rev. D 29 (1984) 1424.

\bibitem{sde} F.J. Dyson, Phys. Rev. 75 (1949) 1736;
J.S. Schwinger, Proc. Nat. Acad. Sci. 37 (1951) 452.

\bibitem{corn01} J.M. Cornwall, Phys. Rev. D 26 (1982) 1453.

\bibitem{corn02} J.M. Cornwall, J. Papavassiliou, Phys. Rev. D 40 (1989) 3474;
J. Papavassiliou, J.M. Cornwall, Phys. Rev. D 44 (1991) 1285.

\bibitem{corn03} D. Binosi, J. Papavassiliou, JHEP 0811 (2008) 063;
D. Binosi, J. Papavassiliou, Phys. Rept. 479 (2009) 1;
A.C. Aguilar, J. Papavassiliou, Phys. Rev. D 81 (2010) 034003.

\bibitem{agpapa} A.C. Aguilar, J. Papavassiliou, Eur. Phys. J. A 35 (2008) 189.

\bibitem{corn04} A.C. Aguilar, J. Papavassiliou, JHEP 0612 (2006) 012.

\bibitem{private} A.C. Aguilar, private communication, 2010.

\bibitem{nat} A.A. Natale, PoS (QCD-TNT 09) 031 (2009); arXiv:0910.5689. 

\bibitem{corn05} J.M. Cornwall, W.S. Hou, Phys. Rev. D 34 (1986) 585.

\bibitem{lav} M. Lavelle, Phys. Rev. D 44 (1991) 26;
D. Dudal, J.A. Gracey, S.P. Sorella, N. Vandersickel, H. Verschelde, Phys. Rev. D 78 (2008) 065047.

\bibitem{valenzuela} G. Cveti\v{c}, R. K\"ogerler, C. Valenzuela, Phys. Rev. D 82 (2010) 114004.

\bibitem{heradata} I. Abt, et al., Nucl. Phys. B 407 (1993) 515;
T. Ahmed, et al., Nucl. Phys. B 439 (1995) 471;
M. Derrick, et al., Z. Phys. C 65 (1995) 379;
M. Derrick, et al., Z. Phys. C 69 (1996) 607;
M. Derrick, et al., Z. Phys. C 72 (1996) 399;
S. Aid, et al., Nucl. Phys. B 470 (1996) 3;
C. Adloff, et al., Nucl. Phys. B 497 (1997) 3;
J. Breitweg, et al., Phys. Lett. B 407 (1997) 432;
J. Breitweg, et al., Eur. Phys. J. C 7 (1999) 609;
J. Breitweg, et al., Phys. Lett. B 487 (2000) 53;
C. Adloff, et al., Eur. Phys. J. C 21 (2001) 33;
S. Chekanov, et al., Eur. Phys. J. C 21 (2001) 443.
 
\bibitem{shirkov01} D.V. Shirkov, I.L. Solovtsov, Phys. Rev. Lett. 79 (1997) 1209.

\end{thebibliography}
\end{document}